\documentclass{article}
\usepackage{spconf}
\usepackage{tabu}

\usepackage[]{siunitx}

\usepackage{cite}
\usepackage{graphicx}
\usepackage[cmex10]{amsmath}
\usepackage{multirow}
\usepackage{makecell}
\usepackage{diagbox}

\usepackage{newtxtext,newtxmath}
\usepackage{bm}
\usepackage{psfrag}
\usepackage{color}
\usepackage{colortbl}
\usepackage{booktabs}

\usepackage{arydshln}

\title{Target speaker voice activity detection with transformers and\\its integration with end-to-end neural diarization}

\makeatletter
\def\@name{\textit{Dongmei Wang, Xiong Xiao, Naoyuki Kanda, Takuya Yoshioka, Jian Wu}\\
}
\makeatother
\address{Microsoft, One Microsoft Way, Redmond, WA, USA}

\begin{document}
\ninept

\maketitle

\begin{abstract}
This paper describes a speaker diarization model based on target speaker voice activity detection (TS-VAD) using transformers. To overcome the original TS-VAD model's drawback of being unable to handle an arbitrary number of speakers, we investigate model architectures that use input tensors with variable-length time and speaker dimensions. Transformer layers are applied to the speaker axis to make the model output insensitive to the order of the speaker profiles provided to the TS-VAD model. Time-wise sequential layers are interspersed between these speaker-wise transformer layers to allow the temporal and cross-speaker correlations of the input speech signal to be captured. We also extend a diarization model based on end-to-end neural diarization with encoder-decoder based attractors (EEND-EDA) by replacing its dot-product-based speaker detection layer with the transformer-based TS-VAD. Experimental results on VoxConverse show that using the transformers for the cross-speaker modeling reduces the diarization error rate (DER) of TS-VAD by 11.3\%, achieving a new state-of-the-art (SOTA) DER of 4.57\%. Also, our extended EEND-EDA reduces DER by 6.9\% on the CALLHOME dataset relative to the original EEND-EDA with a similar model size, achieving a new SOTA DER of 11.18\% under a widely used training data setting. 
\end{abstract}
\section{Introduction}
\label{sec:Introduction}
\vspace{-.3em}

Speaker diarization aims to recognize ``who spoke when'' from multi-talker long-form audio \cite{park_diarization_review_2013}.
It is essential for many 
applications such as automatic meeting note-taking, call center data analytics, and medical dialogue summarization. 
A conventional approach is based on exclusively clustering short-time speech segments into different speakers. 
While being widely used, the clustering-based speaker diarization methods cannot deal with overlapping speech without additional processing steps, such as overlapping speech detection \cite{bullock2020overlap,wang2021bytedance} or speech separation \cite{Wang2021DirHardIII, xiao2021microsoft, Niu2021SepGuideDia}. 
Thus, the overall diarization systems tend to become complex and are hard to optimize end to end.

Recently, several neural network-based speaker diarization methods were proposed to deal with the overlapping speech problem. 
End-to-end neural diarization (EEND) \cite{fujita_eend_interspeech_2019, fujita_eend_asru_2019} directly estimates the voice activities of each speaker with a neural network trained with a permutation-free objective function, also known as permutation-invariant loss, which was originally proposed for speech separation \cite{hershey2016deep,Single_Channel-Isik2016,Yu_pit_2017, kolbak_pit_2017}. 
While the original EEND handles only a fixed number of speakers, 
a newer model called EEND with encoder-decoder based attractors (EEND-EDA) can deal with an unlimited number of speakers~\cite{shota_eendEDA_2020, shota_eendEDA_2022}. This encoder-decoder-based model counts the speakers and generates the speaker representation, called an attractor, for each detected person, which are used to estimate the speaker activities.  
Another way for dealing with the unlimited number of speakers is to combine EEND and clustering~\cite{keisuke_eendVclustering_ICASSP_2021,keisuke_eendVclustering_interspech_2021}.
As of writing, the combination of EEND-EDA and clustering achieves the state-of-the-art (SOTA) diarization error rate (DER) for the CALLHOME evaluation dataset under a fixed training set condition \cite{EDA_global_local}\footnote{For completeness, we note that the best performance was obtained with WavLM-based unsupervised training using a much larger dataset~\cite{chen2021wavlm}.}.

Target-speaker voice activity detection (TS-VAD) is 
another actively studied approach to the speaker-overlap problem~\cite{medennikov2020stc,ivan_TSVAD_2020}, with excellent performance demonstrated in challenging diarization tasks 
such as CHiME-6 \cite{watanabe20b_chime}, DIRHARD-III \cite{Wang2021DirHardIII} and VoxSRC 2021 \cite{duke_VoxSRC_2021, brown2022voxsrc}. 
Given a set of  speaker profiles, such as i-vectors \cite{dehak2010front}, of participating speakers, TS-VAD estimates the voice activities of each speaker. The speaker profiles are usually obtained by performing the clustering-based speaker diarization beforehand. 
The originally proposed TS-VAD model concatenates the internal speaker representations of all the speakers to account for the cross-speaker correlations.  
Therefore, it cannot handle an unlimited number of speakers. 
While ad hoc methods for bypassing this limitation were adopted in prior studies~\cite{duke_VoxSRC_2021, Wang2021DirHardIII}, these methods may limit the overall effectiveness of TS-VAD. 
Very recently, multi-target filter and detector (MTFAD) was proposed to deal with a variable number of speakers in TS-VAD~\cite{chinyi_multiTargetFilter_2022}. However, the model's performance depends on the order with which the speaker profiles are presented\footnote{To be accurate and fair, the manuscript of \cite{chinyi_multiTargetFilter_2022} was updated on arXiv while we were preparing this paper. Although the updated manuscript mentions the potential use of transformers, no experimental results are reported.}.

In this paper, we examine the use of transformers \cite{transformer_2017} for the cross-speaker modeling in TS-VAD to cope with any number of speakers while keeping the model's output invariant to the speaker profile order. 
To this end, we use an input tensor with variable-length time and speaker dimensions. 
The transformer layers are applied to the speaker axis without positional encoding to make the model order-invariant with respect to the speaker profiles. 
Another set of sequential layers, such as bidirectional long short-term memory (BLSTM) or transformer layers, are interspersed between the speaker-wise transformer layers to help the model capture the correlations both in time and across the speakers. 
Furthermore, we show that the transformer-based TS-VAD can extend EEND-EDA by replacing its dot-product-based speaker detection layer. 
A comprehensive evaluation using the VoxConverse dataset~\cite{Voxconverse2021dataset} shows the effectiveness of  the transformer-based cross-speaker modeling in TS-VAD.
The proposed EEND-EDA model enhanced with TS-VAD also significantly reduces the DER on the CALLHOME dataset~\cite{callhome}.

\section{Related works}
\vspace{-.3em}

\subsection{TS-VAD}
\vspace{-.3em}

TS-VAD was proposed to detect the speech activities of a set of target speakers in the input speech signal based on the target speakers' profiles (e.g. i-vectors)~\cite{medennikov2020stc,ivan_TSVAD_2020}.
The TS-VAD model consists of three modules: 1) a convolutional neural network (CNN) based encoder that extracts embedding vectors from the input signal, 2) a BLSTM-based independent-speaker-detection (ISD) module that processes each target speaker independently based on the speaker's profile and the embedding vectors, 3) a BLSTM-based joint-speaker-detection (JSD) module that models both cross-time and cross-speaker information and predicts the activities of all speakers simultaneously. 
The model is trained to minimize the sum of the binary cross-entropy losses for all the target speakers. 
Since the joint modeling in the JSD module is realized by concatenating the ISD model outputs for all speakers, the input dimension of the BLSTM in the JSD module is determined by the number of target speakers. Hence, the maximum number of speaker is built into the model, making it inflexible in handling an arbitrary number of speakers.

\begin{figure}[t]
    \centering
    \includegraphics[scale=0.55]{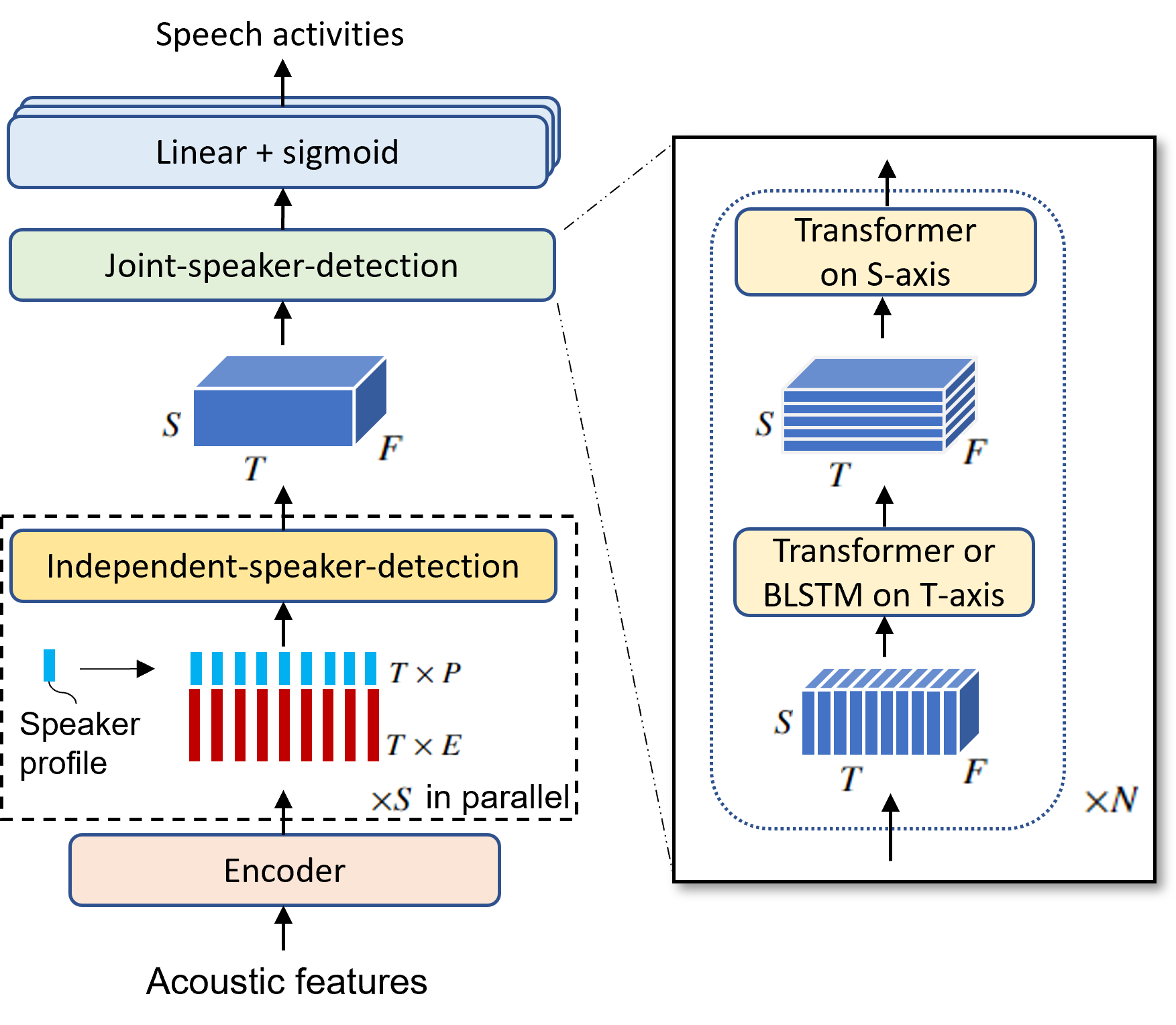}
    \vspace{-1em}
    \caption{Architecture of transformer-based TS-VAD. BLSTM and transformer are applied through the sliced dimension.}
    
    \vspace{-1.5em}
    \label{fig: U-TS-VAD}
\end{figure}

\subsection{EEND-EDA}
\label{sec:eend-eda}
\vspace{-.3em}

Unlike TS-VAD, which usually relies on another diarization method to provide profiles, EEND-EDA is an end-to-end speaker diarization model for an unlimited number of speakers and does not require a separate clustering step~\cite{shota_eendEDA_2020, shota_eendEDA_2022}.
It consists of 
a transformer-based encoder module, 
an encoder-decoder-based attractor (EDA) 
extraction module, and a matching module. 
The encoder module generates 
frame embeddings from the input speech signal.
The EDA extraction module estimates the number of speakers in the input signal and produces an attractor to represent each of them. Specifically, it first scans through the embedding sequence with an LSTM-based encoder, and then produces speaker attractors one by one using an LSTM-based decoder. 
Each attractor is supposed to represent one particular speaker in the input audio and has an attractor existence probability.
The decoding stops when the estimated existence probability of the last attrator becomes below a pre-defined threshold. 
Finally, the matching module estimates the speaker activities by calculating  
 the dot-product between the frame embeddings from the encoder module
 and the speaker attractors from the EDA module. 
 The EEND-EDA model is trained to minimize the weighted sum of
 a permutation-free diarization loss and a speaker existence loss.    
Note that the functionality of the matching module is the same as that of TS-VAD as both models estimate the speech activities based on a set of speaker attractors or profiles.

\begin{figure}[t]
    \centering
    \includegraphics[scale=0.56]{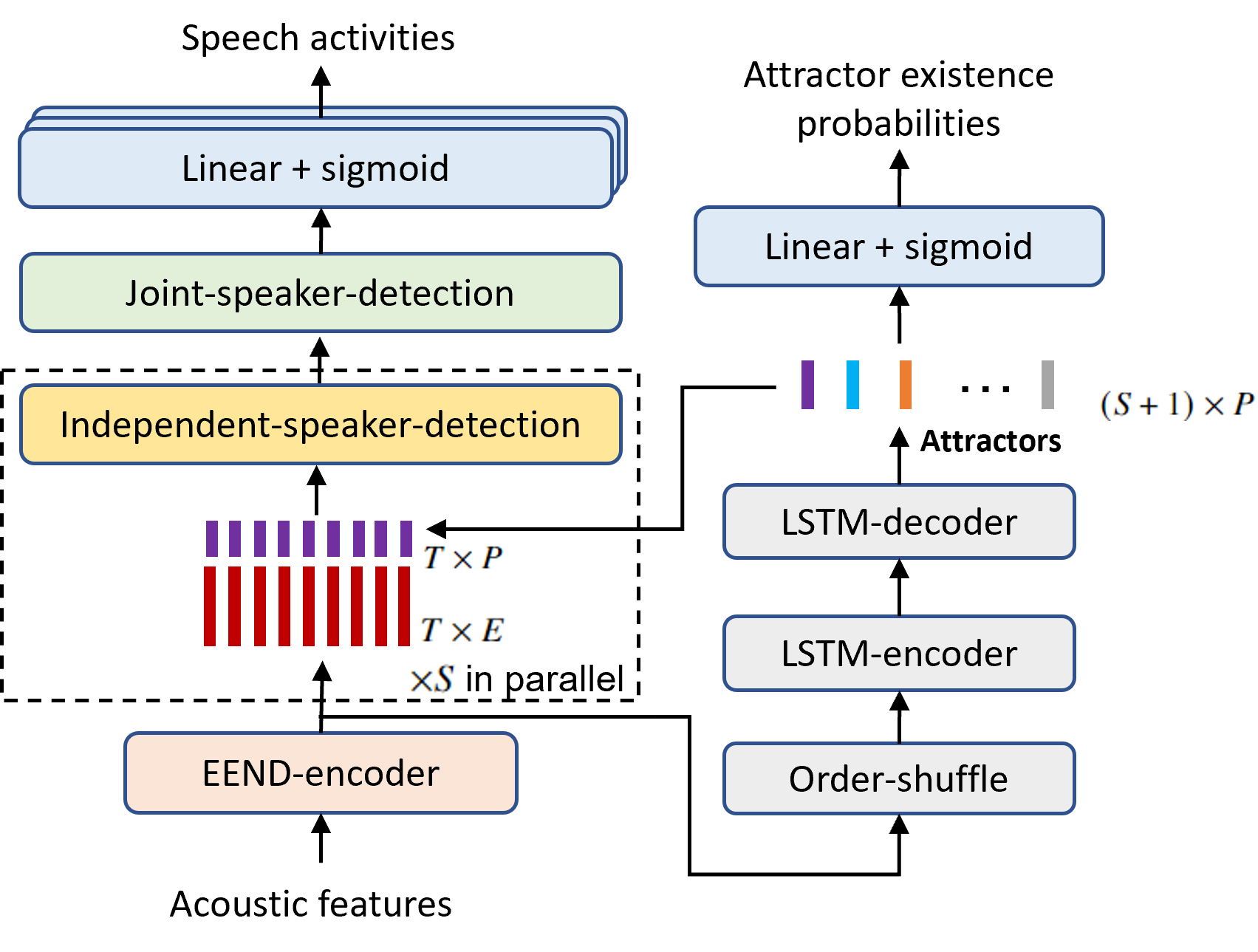}
    \vspace{-1em}
    \caption{EDA-TS-VAD architecture.} 
    \vspace{-1.5em}
    \label{fig: eda-TS-VAD} 
\end{figure}

We speculate that the use of the dot-product in the matching module may limit EEND-EDA's effectiveness.
For an audio segment containing overlapping speech, the dot-product between the frame embeddings and the speaker attractors must have large values
for all the attractors of the active speakers. 
 However, it would be difficult to satisfy this requirement 
because the dot-product of two vectors essentially measures 
the proximity between the vector directions while different attractors should be pointing at different directions. 

\section{TS-VAD with Transformer}
\vspace{-.3em}

One of our contributions is to perform a comprehensive investigation of the effectiveness of 
TS-VAD models with transformer layers applied for cross-speaker modeling. 
These models can deal with an arbitrary number of speaker profiles. 

Fig.~\ref{fig: U-TS-VAD} shows the examined model architecture, which consists of the encoder, ISD, and JSD modules following the original TS-VAD~\cite{medennikov2020stc,ivan_TSVAD_2020}.
The encoder module takes in log mel-filterbank features
and outputs the frame embeddings $\textbf{E}\in\mathbb{R}^{T\times E}$, where $T$ and $E$ are
the number (i.e., the sequence length) and dimension of the embeddings, respectively.
The ISD module processes each target speaker independently based on their profiles. The profiles are denoted as $\textbf{P}\in\mathbb{R}^{S\times P}$, where $S$ is the number of the target speakers and $P$ is
the profile dimension. 
Specifically, for the $i$-th target speaker, the input to the ISD module has a shape of $\mathbb{R}^{T\times (E+P)}$, which is obtained by appending row vector $\textbf{p}_i$ to all the $T$ embedding vectors in $\textbf{E}$. 
The ISD module
consists of a linear projection layer, followed 
by two BLSTM layers.
The output from the ISD
module is $S$ matrices of shape $\mathbb{R}^{T\times F}$.

Unlike the original TS-VAD which concatenates these $S$ speaker-specific frame embeddings to form a two-dimensional tensor of shape $\mathbb{R}^{T\times (S\cdot F)}$,
we regard it
as a three-dimensional tensor of shape
$\mathbb{R}^{T\times S \times F}$.
The sequence modeling layers are then applied along $S$-axis and $T$-axis alternately. 
We adopt the transformer without positional encoding for $S$-axis to model the cross-speaker correlation so that the model's output becomes independent of the order of the speaker profiles. For $T$-axis, we examine both BLSTM and transformer to model the temporal correlation\footnote{In our experiments, we also consider using BLSTM for both $S$-axis and $T$-axis as it is equivalent to MTFAD~\cite{chinyi_multiTargetFilter_2022}. Note that the resultant model's performance becomes dependent on the speaker profiles' order.}.
After the JSD layer, we apply a linear layer with sigmoid activation to obtain the speech activities for all target speakers.
The model is trained to minimize the sum of the binary cross-entropy losses of all speakers.

\section{Integration of TS-VAD with EEND-EDA}
\vspace{-.3em}

As another major contribution, 
we also integrate the transformer based TS-VAD into the EEND-EDA \cite{shota_eendEDA_2020} to overcome the latter's limitation discussed in Section \ref{sec:eend-eda}.
Considering the shared functionality of EEND-EDA's matching module and TS-VAD as we noted earlier and TS-VAD's ability to handle speech overlaps,
we propose to replace the dot product based matching module with the TS-VAD (excluding the encoder module).
We call this model EDA-TS-VAD. 

Fig.~\ref{fig: eda-TS-VAD} shows the model architecture of EDA-TS-VAD. 
Following the EEND-EDA framework, EDA-TS-VAD includes a transformer 
based encoder module and 
an LSTM-based EDA module, which generate the frame embeddings and speaker attractors, respectively. 
The embeddings and attractors are then 
fed to the TS-VAD-based matching module to estimate the speech activities of each speaker. 
Note that TS-VAD can be used here because the transformer-based TS-VAD model can deal with any number of speakers and its outputs are insensitive to the attractor order. 
The model is trained with the same loss as EEND-EDA.

\section{Experimental results}
\vspace{-.3em}

We report our experimental results for the transformer-based TS-VAD and EDA-TS-VAD in Sections
\ref{sec:tsvad_exp} and \ref{sec:eda_exp}, respectively.

\subsection{TS-VAD evaluation with VoxConverse test set}
\label{sec:tsvad_exp}
\vspace{-.3em}

\subsubsection{VoxConverse test set}
\vspace{-.3em}
The effectiveness of the TS-VAD models was evaluated by using the VoxConverse dataset \cite{Voxconverse2021dataset}. 
It was originally developed based on YouTube videos for audio-visual
diarization and used for the audio-based diarization track of VoxSRC 2020~\cite{nagrani2020voxsrc} and VoxSRC 2021 challenges~\cite{brown2022voxsrc}.
We used the dev set (216 sessions, 20.3 hours, 1-20 speakers in each session) to fine-tune the models, and the test set (232 sessions, 43.5 hours, 1-21 speakers in each session) 
for evaluation. 
The DER was calculated with a 0.25-sec collar tolerance \cite{Voxconverse2021dataset}.

\subsubsection{Configuration of TS-VAD systems}
\vspace{-.3em}

To extract speaker profiles, we used a 17-layer ResNet-based d-vector extractor \cite{zhou2019cnn} without the final averaging layer. The input to the d-vector model was a sequence of 80-dim log mel-filterbank features extracted every 10 msec, and the output was one 128-dim d-vector per 80 msec. The d-vector extractor was trained by using
the VoxCeleb1 dev set~\cite{nagrani17VoxCeleb} and the VoxCeleb2 dev and test sets~\cite{Chung18b}.

The TS-VAD encoder module was configured to be exactly the same as the d-vector extractor. 
The ISD module consists of a linear projection layer followed by two BLSTM layers. 
Finally,
the JSD module consists of two blocks of sequence modeling layers, with each block consisting of speaker-axis and time-axis sequence layers as shown in Fig.~\ref{fig: U-TS-VAD}.
We tested transformer and BLSTM for these sequence modeling layers.
The number of hidden units in the transformer and BLSTM layers are configured such that the variants of TS-VAD have similar numbers of parameters, ranging from 7.8M to 8.4M. Refer to Appendix for the details of the model configuration.

The TS-VAD models were first pre-trained with simulated conversations based on the VoxCeleb1 dev set \cite{nagrani17VoxCeleb} and the VoxCeleb2 dev and test sets \cite{Chung18b}. 
It was then fine-tuned by using the VoxConverse dev set.
At the begining of the training, the parameters of the TS-VAD were randomly initialized except that the encoder module was initialized with the parameters of the d-vector extractor. During training, for each speaker, we obtained the speaker profile by taking the mean of the d-vectors belonging to the target speaker while excluding the overlapping speech regions. 
The simulated conversations were generated by combining source sentences of 2-10 speakers from VoxCeleb1 and VoxCeleb2 with  overlap ratios of up to 30\%. Additive background noise and reverberation from various sources were then applied to the simulated conversations. In total, 539k simulated conversations of about 19.5k hours were generated, with each conversation's duration varying from 13 sec to about 8 min. 
These conversations were chopped up into 60-sec long training samples. 
The models were trained with a batch size of 32. Linear decay learning rate scheduler with warm-up was used, with a peak learning rate of 2e-4. The numbers of the warm-up and total training steps were 20k and 200k, respectively. 
After the simulation-based pre-training, 
the model was fine-tuned by using the VoxConverse dev set, where 
a fixed learning rate of 1e-5 was used.

\subsubsection{Inference settings}
\vspace{-.3em}

In the inference, the test audios were first diarized with the agglomerative hierarchical clustering (AHC) method of \cite{xiao2021microsoft} by using a Res2Net d-vector model \cite{zhou2021resnext}. 
The speaker profile was then extracted from all the detected speech regions for each speaker.
For speakers whose detected speech segments were shorter than 2 sec in total, no profiles were created and their speech activities were copied to the final diarization results. 
The test audios were chopped up into 60-sec chunks to match the training segment length. An 11-tap median filter was applied on the frame-level diarization decisions. 

For the baseline, we trained an original TS-VAD model with 10 speaker profiles by using the same training configuration.
If the number of speakers detected by AHC was smaller than 10, 
zero vectors were appended to make the number of speaker profiles 10. 
When AHC detected more than 10 speakers, the speaker profiles of the top-10 speakers ranked by their active speech durations were fed to the TS-VAD model, while the speech activities of the excluded speakers were copied from the AHC to the final output~\cite{duke_VoxSRC_2021, Wang2021DirHardIII}.

\begin{table}[t]
\tabcolsep = 1.3mm
\centering
\caption{DER (\%) on VoxConverse test set with 0.25 sec of tolerance collar. Systems S4 and S5 correspond to the original TS-VAD and MTFAD, respectively. Version 0.3 reference was used except for S2. }
\label{tab:VoxSRC-result}
  \vspace{0.5mm}
{\footnotesize
\begin{tabular}{@{}llcccrrr@{}}
\toprule
\multirow{2}{*}{ID} & \multirow{2}{*}{System} &  \multicolumn{2}{c}{Joint-speaker-detection}    & \multirow{2}{*}{Profiles} & \multicolumn{3}{c}{DER (\%)} \\ \cmidrule{3-4} \cmidrule{6-8}
        &        &  Time & Speaker     &    & 1-10 & 11+ & Total \\
\midrule
S1 & Xiao et al. \cite{xiao2021microsoft}$^\dagger$ & - & - & -   & 5.47 & 8.45    & 6.08                        \\
S2 & Wang et al. \cite{wang2021bytedance}$^\ddagger$ & - & - & -  & - & -    & 5.17                        \\
S3 & AHC      & - & -      & -     & 6.61 & 9.27  & 7.15                        \\
\midrule
S4 & \multirow{6}{*}{TS-VAD}  & BLSTM & Concat.     & S3  & 4.68 & 7.79   & 5.31                        \\
S5 &     & BLSTM & BLSTM     &  S3  & 4.44 & 6.74  & 4.91                        \\
\cmidrule{3-8}
S6 &     & Trans. & Trans.     & S3  & 4.32 & 6.64 & 4.80                        \\
S7 &     & BLSTM & Trans.    &  S3  & 4.26 & 6.47  & 4.71                        \\
S8 &     & BLSTM & Trans.    &  S1  & 4.10 & 6.37 & 4.57                        \\ 
S9 &     & BLSTM & Trans.    & Oracle  & 3.21 & 4.35 & 3.45                       \\
\bottomrule
\end{tabular}
}
\\{\scriptsize \hspace{-26mm}$^\dagger$ VoxSRC 2020 1st-ranked system with
5-system fusion. 
\\$^\ddagger$ VoxSRC 2021 2nd-ranked system with 3-system fusion. Note that VoxSRC 2021\\
\vspace{-0.5mm}\hspace{-16mm}1st-ranked system did not report the DER for this test subset.}
\vspace{-6mm}
\end{table}

\begin{table*}[t]
\centering
\caption{DER (\%) on CALLHOME dataset with 0.25 sec of colloar tolerance. $^\star$ Provided by the authors of \cite{shota_eendEDA_2022,EDA_global_local}.}
\label{tab:der-edaTSVAD}
{\footnotesize
\begin{tabular}{lcccccccccc}
\toprule
\multirow{2}{*}{ID} &
  \multirow{2}{*}{Models} &
  \multirow{2}{*}{\begin{tabular}[c]{@{}c@{}} \#Transformers in\\EEND-encoder\end{tabular}} &
  \multirow{2}{*}{\begin{tabular}[c]{@{}c@{}}\#JSD\\ -blocks\end{tabular}} &
  \multirow{2}{*}{\begin{tabular}[c]{@{}c@{}}Model\\ size (M)\end{tabular}} &
  \multicolumn{5}{c}{\#Speakers} &
  \multirow{2}{*}{Total} \\ \cmidrule{6-10}
                                 &   &   &   &        & 2    & 3     & 4     & 5              & 6              &                \\ 
\midrule 
B1 & EEND-EDA \cite{shota_eendEDA_2022}                         & 4  & - & \hspace{3mm}6.4$^\star$     & \hspace{2mm}8.09$^\star$    & \hspace{1mm} 12.20$^\star$    & \hspace{1mm} 15.32$^\star$      &\hspace{1mm} 27.36$^\star$          & \hspace{1mm} 29.21$^\star$            & 12.88          \\
B2 & EEND-EDA+clustering \cite{EDA_global_local}                & 6  & - & \hspace{2mm}10.7$^\star$    & 7.11    & 11.88    & 14.37      & 25.95          & \textbf{21.95}   & 11.84         \\ 
\hdashline[1pt/2pt]\hdashline[0pt/1pt]  
B3 & EEND-EDA (our impl.)                           & 6  & - & 12.2 & \textbf{6.82} & 11.06 & 16.45 & 22.62          & 32.15          & 12.23          \\
B4 &  EEND-EDA (our impl.)                                                         & 8 & - & 15.6 & 6.84 & 11.02 & 16.08 & \textbf{21.73} & 31.99          & 12.01          \\
 \hdashline[1pt/2pt]\hdashline[0pt/1pt] 
P1 & EDA-TS-VAD                                & 6  & 1 & 14.7 & 7.04 & \textbf{10.39} & 15.38 & 25.48          & 31.41          & 11.91          \\
P2 &   EDA-TS-VAD                                                        & 6  & 2 & 16.1 & 7.33 & 10.50 & \textbf{12.97} & 22.65          & 24.48 & \textbf{11.18} \\ 
\bottomrule
\end{tabular}
}
\vspace{-6mm}
\end{table*}

\subsubsection{Results}
\vspace{-.3em}

Table~\ref{tab:VoxSRC-result} shows the DERs of TS-VAD with various model architectures as well as  
several reference systems. 
By comparing S4 (original TS-VAD) with S6 and S7, we can observe consistent DER improvement by the TS-VAD with transformer based cross-speaker modeling. 
The degree of improvement was greater for the subset with more than 10 speakers (shown as ``11+'').
For example, S7 outperformed S4 by 16.9\% relative for the 11+ speaker subset while the relative DER gain was 9.0\% for the 1-10 speaker subset.
This result validates our hypothesis that the heuristics used for the original TS-VAD \cite{duke_VoxSRC_2021, Wang2021DirHardIII} limits
the diarization accuracy. 

Note that S5 is equivalent to MTFAD~\cite{chinyi_multiTargetFilter_2022} , which uses BLSTM for both the time and speaker axes.
Comparison of S5 and S7 reveals the benefit of using the transformer instead of BLSTM for modeling the cross-speaker correlations
as the latter does not show dependency on the speaker profile order. 
On the other hand, comparing S6  with S7 shows that using BLSTM for the time axis slightly outperformed the system using the transformer for the time axis. This might be because our transformer did not use positional encoding, which could be useful for time sequence modeling.

Finally, from the results of S7-S9,  we can observe a positive correlation between the speaker profile quality and the final diarization accuracy. 
The TS-VAD model with transformer and BLSTM for the speaker- and time-axis modeling, respectively, that uses S1 as the first diarization step achieved a SOTA DER of 4.57\%.

\subsection{EDA-TS-VAD evaluation with CALLHOME test set}
\label{sec:eda_exp}
\vspace{-.3em}

\subsubsection{CALLHOME}
\vspace{-.3em}

The CALLHOME dataset \cite{callhome} consists of  conversational telephone speech recordings. 
We followed the data split used in \cite{shota_eendEDA_2020}, 
where the Part 1 subset 
contains 249 sessions (8.70 hours; 2 to 7 speakers in each session) 
and the Part 2 subset contains 250 sessions (8.55 hours; 2 to 6 speakers in each session). 
Part 1 was used for model fine-tuning. Part 2 was used for the evaluation.
The DER evaluation used a 0.25-sec tolerance collar.

\subsubsection{Configuration of EDA-TS-VAD}
\vspace{-.3em}

The EDA-TS-VAD model takes a 600-dim feature vector as an input, which is
created by stacking 15 frames of 40-dim log mel-filterbank features.  
The input frames are subsampled by a factor of 10.
EEND-encoder has a linear layer projecting the feature dimension from 600 to 320, followed by 6 stacked transformer encoder layers with 10 attention heads and 320-dim query/key/value hidden embeddings in each self-attention
module.  
The encoder and decoder of the EDA consist of one unidirectional LSTM layer with a hidden size of 320. By following the setup of \cite{shota_eendEDA_2020}, the final hidden and cell states of the encoder are used to initialize the decoder. 
A linear layer followed by sigmoid function is applied to each attractor to estimate the attractor existence probability, or the probability that each attractor corresponds to one of the speakers participating in the input speech. 
The ISD module consists of the linear layer followed by 2 
BLSTM layers. 
The JSD module uses transformer for the speaker axis and
BLSTM for the time axis.

In the training, we first pre-trained the model based on simulated mixtures and then fine-tuned it by using the CALLHOME Part 1 subset.
For the speech mixture generation, 
 we followed the data simulation protocol proposed in \cite{fujita_eend_interspeech_2019} to create $2/3/4$-speaker mixtures.
 This was done by using the speech samples taken from Switchboard-2, Switchboard Cellular, 
and NIST Speaker Recognition Evaluation
and  the noise samples taken from MUSAN corpus \cite{musan_corpus}. 
In the pre-training, 
the EEND-encoder module is initialized with the corresponding part of a well-trained EEND-vector-clustering model \cite{keisuke_eendVclustering_ICASSP_2021}. We used the same training data for both the EEND-vector-clustering and EDA-TS-VAD models. 
After the initialization, the entire model was trained 
with a linearly decaying learning rate after a linear warm-up period.
The peak learning rate was set to 5e-5, the batch size was 36, and
 the number of the warm-up and overall training steps were set to 27k and 230k steps (roughly corresponding to $50$ epochs), respectively. 
For fine-tuning, 
the model was trained for 100 epochs with a batch size of 4 using CALLHOME Part1. 
Linear decay learning rate scheduler is used with a peak learning rate of 2e-5. 
The data was chunked into 50 sec and 200 sec segments for processing in the pre-training and fine-tuning stages, respectively.
During the model training,
we shuffled the embeddings across time frames before feeding them to the EDA module as proposed in \cite{shota_eendEDA_2020}.

\subsubsection{Results}
\vspace{-.3em}

Table~\ref{tab:der-edaTSVAD} shows the DER results of the proposed EDA-TS-VAD and various baseline systems.
Comparing B3, P1, and P2 shows that integrating TS-VAD  into EEND-EDA consistently improved the DER compared with the original EEND-EDA. 
For fair comparison with respect to the model size, we also trained a large EEND-EDA by increasing the number of transformer layers in the EEND-encoder module (B4). 
This yielded an only marginal DER improvement,
suggesting that the dot-product operation 
is the performance bottleneck in the original EEND-EDA.
The proposed EDA-TS-VAD (P2) achieved a DER of 11.18\%, or a 
relative DER reduction of 6.9\%  
compared with the EEND-EDA with a similar model size (B4).

\vspace{-.3em}


\section{Conclusion}
\vspace{-.3em}

We investigated TS-VAD models using transformers for cross-speaker correlation modeling in speaker diarization. These models can handle an arbitrary number of speakers in such a way that the model output is invariant to the order of the speaker profiles. 
A significant DER reduction was achieved on the VoxConverse test set, especially when the speaker number was large. 
We further integrated this transformer based TS-VAD into 
EEND-EDA by replacing the dot-product operation to deal with overlapped speech more effectively.
Our evaluation on the CALLHOME data set showed the effectiveness of the proposed model integration.

\section{Acknowledgement}
\vspace{-.3em}
We thank Shota Horiguchi of Hitachi for sharing and allowing us to include the detailed results of their previous work~\cite{shota_eendEDA_2022,EDA_global_local}. We thank Yong Zhao of Microsoft for sharing the ResNet-based d-vector extractor used for the speaker profile extraction.

\bibliographystyle{IEEEbib}
\bibliography{my_references,refs}

\newpage
\appendix

\section{TS-VAD Configuration in VoxConverse evaluation}
\label{appendix}

In this appendix, 
we describe the detailed model configurations of the TS-VAD variants used in Table~\ref{tab:VoxSRC-result}.
All the variants share the same encoder and ISD module configuration. The encoder is a 17-layer ResNet-based d-vector extractor~\cite{zhou2019cnn}. In the ISD module, a linear layer first projects 256-dim features to 384-dim features, then 2 layers of BLSTM are applied. Each BLSTM layer contains 128 memory cells in each direction, followed by a linear transformation layer from 256-dim to 256-dim.

There are four JSD variants. Table~\ref{tab:config_jsd} shows
their details. While the original TS-VAD (S4) has only one BLSTM layer in the JSD module, the other variants have two blocks of sequence modeling layers. In each block, there is a sequence modeling layer for the time axis, followed by a sequence modeling layer for the speaker axis.  In System S4, the maximum number of target speakers is set to 10. 

\begin{table}[h]
\centering
\caption{Configurations of JSD variants of TS-VAD. $\text{BLSTM}(x,y,z)$ means BLSTM layer with $x$ being the input dimension, $y$ the number of memory cells in each direction, and $z$ the number of projected dimensions. $\text{Transformer}(p, q, t)$ means transformer layer with $p$ being the number of attention heads, $q$ the dimension of embedding in the self-attention layer, and $t$ the dimension of the position-wise feed-forward layer.}
\label{tab:config_jsd}
 \vspace{0.5mm}
\begin{tabular}{@{}lllll@{}}
\toprule 
\multirow{2}{*}{ID}  &  \multicolumn{2}{c}{Joint-speaker-detection module} \\ 
\cmidrule{2-5}
           &  \multicolumn{1}{c}{Time axis model} & \multicolumn{1}{c}{Speaker axis model}\\
\midrule
S4    & BLSTM(2560, 256, 192)          & Concatenation           \\
S5    & BLSTM(256, 160, 160)          & BLSTM(256, 160, 160)          \\
S6    & Transformer(4, 256, 256)       & Transformer(4, 256, 256) \\
S7-S9 & BLSTM(256, 160, 160)          & Transformer(4, 160, 160) \\
\bottomrule
\end{tabular}
\end{table}

\end{document}